\documentclass[copyright,creativecommons]{eptcs}
\providecommand{\event}{GANDALF 2014} 
\newif\ifeptcs\eptcstrue
\newif\iflncs\lncsfalse

\iflncs
\else
\usepackage{amsthm}
\fi

\usepackage[numbers,sort]{natbib}
\usepackage{url}
\usepackage{fixme}
\usepackage{xspace}
\usepackage{verbatim}
\usepackage{amssymb}
\usepackage{amsmath} 
\usepackage{amsfonts} 
\usepackage[mathscr]{eucal}
\usepackage{listings}
\usepackage{paralist}

\iflncs
\spnewtheorem{exmp}{Example}[section]{\bfseries}{\rmfamily}
\spnewtheorem{alg}{Algorithm}[section]{\bfseries}{\rmfamily}
\spnewtheorem{corol}{Corollary}[section]{}{}
\else
\newtheorem{theorem}{THEOREM}[section]

\newtheorem{corol}[theorem]{COROLLARY}
\newtheorem{lemma}[theorem]{LEMMA}

\theoremstyle{definition}
\newtheorem{definition}[theorem]{DEFINITION}

\newtheorem{example}[theorem]{EXAMPLE}

\fi

\let\com=\newcommand
\com{\bthm}{\begin{theorem}}
\com{\ethm}{\end{theorem}}
\com{\bdfn}{\begin{definition}}
\com{\edfn}{\end{definition}}
\com{\blem}{\begin{lemma}}
\com{\elem}{\end{lemma}}
\com{\bcor}{\begin{corol}}
\com{\ecor}{\end{corol}}
\com{\bexm}{\begin{example}}
\com{\eexm}{\end{example}}
\com{\bprf}{\begin{proof}}
\iflncs
\com{\eprf}{\qed\end{proof}}
\else
\com{\eprf}{\end{proof}}
\fi

\newcommand{\trans}[0]{{\mathrm{T}}}

\newcommand{\tuple}[1]{\langle #1 \rangle}

\newcommand{\ints}{\ensuremath{\mathbb Z}\xspace}
\newcommand{\nats}{\ensuremath{\mathbb N}\xspace}
\newcommand{\rats}{\ensuremath{\mathbb Q}\xspace}

\let\vect=\vec
\renewcommand{\vec}[1]{\mathbf{#1}}

\newcommand{\lrf}[0]{\ensuremath{\mathit{LRF}}\xspace}
\newcommand{\lrfs}[0]{{\textit{LRF}s}\xspace}

\newcommand{\linrfp}[0]{\ensuremath{\textsc{LinRF}^{\rho}}\xspace}
\newcommand{\linrfqp}[0]{\ensuremath{\textsc{LinRF}^{\rho}(\ensuremath{\rats})}\xspace}
\newcommand{\linrfzp}[0]{\ensuremath{\textsc{LinRF}^{\rho}(\ensuremath{\ints})}\xspace}

\newcommand{\slcp}[0]{\ensuremath{\mathit{SLC}^{\rho}}\xspace}
\newcommand{\poly}[1]{{\mathscr #1}}

\newcommand{\trcv}[2]{\ensuremath{\bigl(\begin{smallmatrix}{#1}\hfill\\{#2}\end{smallmatrix}\bigr)}}
\let\tr=\trcv
\newcommand{\rfcoeff}[0]{\lambda}

\newcommand{\inv}[0]{\textit{INV}\xspace}
\newcommand{\states}[0]{\ensuremath{\poly{S}}\xspace}
\newcommand{\initials}{\poly{I}}
\newcommand{\transitions}{\poly{Q}}
\let\stopsign=\Box

\newenvironment{fig0}[3]
{
\xdef\fighack{\noexpand\caption{#1}\noexpand\label{#3}}
\begin{figure*}[#2]
\begin{minipage}{0.95\textwidth}}{\fighack\end{minipage}%
\end{figure*}}

\begin{document}

\title{The Hardness of Finding Linear Ranking Functions for Lasso Programs}

\ifeptcs
\author{Amir M. Ben-Amram
\institute{The Academic College of Tel-Aviv Yaffo}
\email{amirben@cs.mta.ac.il}
}
\def\titlerunning{Linear Ranking Functions for Lasso Programs}
\def\authorrunning{A.M. Ben-Amram}
\else
\author{Amir M. Ben-Amram\thanks{The Academic College of Tel-Aviv Yaffo}}
\fi

\maketitle

\begin{abstract}
Finding whether a linear-constraint loop has a linear ranking function is an important key to understanding
the loop behavior, proving its termination and establishing iteration bounds. If no preconditions are provided,
the decision problem is known to be in coNP when variables range over the integers and in PTIME for the rational numbers,
or real numbers.
Here we show that deciding whether a linear-constraint loop with a precondition, specifically with partially-%
specified input,
has a linear ranking function is
EXPSPACE-hard over the integers, and PSPACE-hard over the rationals. The precise complexity of these decision problems is yet unknown.
The EXPSPACE lower bound is derived from the reachability problem for Petri nets (equivalently, Vector
Addition Systems), and possibly indicates an  even stronger lower bound (subject to open problems in VAS theory).
The lower bound for the rationals follows from a novel simulation of Boolean programs.  Lower bounds are also
given for the problem of deciding if a linear ranking-function supported by a particular form of inductive invariant
exists. For loops over integers, the problem is PSPACE-hard for convex polyhedral invariants and EXPSPACE-hard for
downward-closed sets of natural numbers as invariants.
\end{abstract}

\section{Introduction}

\begin{fig0}{A loop with a stem (a): the stem is the straight-line code preceding the {while} loop.
The loop has the ranking function $X$, but this is only justified when the stem is taken into account.
In (b), the loop is written in the formalism of linear constraints.}{t}{fig-simplexample} 
\begin{tabular}{c@{\hspace{0.2\textwidth}}c}
\parbox{0.3\textwidth}{
\begin{tabbing}
$Y := 2$; \\
$\mathit{while}$\=~ $ X > 0 ~\mathit{do}$ \\
\> $Y := Y-1$;  \\
\> $X := X+Y$;  \\
\> $Y := 2*Y$;
\end{tabbing}
} &
\parbox{0.5\textwidth}{
\noindent
$Y = 2$; \\
$\mathit{while}~ X > 0 ~\mathit{do}$\\
\kern 1cm $X' = X+Y-1,\ \ Y' = 2Y-2$\\
\ \\   
\ 
}\\
(a) & (b) \\
\end{tabular} 
\end{fig0}

The results in this paper relate two basic problems in the analysis of loops: reachability and the existence of a linear ranking function
that proves termination of the loop.    We only consider the (often used) model in which loops compute over numeric variables
(most frequently integer) and their effect is expressed by linear equations or inequalities (constraints).

Termination provers, of which \textsc{Terminator} by Cook, Podelski and Rybalchenko~\cite{CPR06}
is a prototypical example,
are based on the subproblem of proving termination for simple loops with
a ``stem", the so-called \emph{lasso} (Figure~\ref{fig-simplexample}).
Termination of such loops is established in \textsc{Terminator} by 
abstracting the loop to linear constraint form and finding a 
 linear ranking function (a function of the state variables which is bounded below and decreases in every iteration). But 
the algorithm used in \textsc{Terminator} to check for the existence of such a 
function~\cite{DBLP:conf/vmcai/PodelskiR04}
does not take the effect of the
``stem," which is a precondition for the simple loop, into account. We may describe the problem solved by such an algorithm as
finding a \emph{universal} ranking function---one that works for any initial state.

There are several works that do take preconditions into account in the algorithm that looks for 
ranking functions. 
Early approaches~\cite{SohnVanGelder:91,CS:01}
used precomputed invariants, and once these invariants were included
in the description of the loop, looked for a universal ranking function.  
Later, some works attempted to integrate the discovery
of \emph{supporting invariants} with the search for a ranking function, e.g., \cite{BMS:LexLinear,HeizmannHLP:ATVA2013}.
Other works heuristically find some precondition under which a ranking function can be established, e.g., 
\cite{cookGLRS2008}.

I am aware of no published upper or lower bounds on the complexity of precisely answering the question:
given a linear-constraint loop with a precondition, does it have a linear ranking function?
This contrasts with the well-understood classification of the 
universal linear ranking-function problem: as a decision problem (for simplicity we only consider decision problems
when referring to a complexity class) it is PTIME over the rationals%
\footnote{We say that we solve the problem over the rationals when we consider the state space to consist
of all rational-valued points that satisfy the loop constraints, and ``over the integers" when only integer points are
considered. See Section~\ref{sec:prelim}.}%
~\cite{DBLP:conf/vmcai/PodelskiR04,DBLP:journals/iandc/BagnaraMPZ12}
and coNP-complete over the integers~\cite{Ben-AmramG13jv}.

In this paper, we show that deciding whether a linear-constraint loop with a precondition of a simple form
has a linear ranking function is EXPSPACE-hard over the integers, and PSPACE-hard over the rationals.  Clearly, these 
problems are much harder than the universal linear ranking-function problem. In fact, we do not even know
if they are decidable!

A possible reaction to the hardness of this problem is to look at a mitigated problem that has been attempted by work already mentioned:
the \emph{invariant-supported ranking function}.  Instead of asking for a ranking function that holds for the precise set of reachable
states, we relax the requirement so that the ranking function has to hold in a set that contains the reachable states, a loop invariant.
Moreover, we consider \emph{inductive invariants}: such an invariant is verified by a local condition, that is, a condition on a single
loop step, and this condition becomes clearly decidable if the invariant comes from a suitable ``effective" class.
  We shall consider two classes of invariants which seem natural: (1) convex polyhedra, that is, conjunctions of linear constraints,
  and (2) disjunctive invariants of a very simple form
(downward-closed sets---basically a union of boxes with one corner at the origin).
 For precise definitions see the Section~\ref{sec:prelim}.
Do they make the problem more tractable? We do not know exactly. But we can show that---over the integers, at least---the problem
is certainly not \emph{easy}. We prove PSPACE-hardness for ranking functions supported by inductive invariants which 
are convex polyhedra, and EXPSPACE-hardness for downward-closed sets.

Thus the results of this paper are four hardness results: two for the general problem with a precondition and two for the invariant-supported
problem.   In addition, for the integer case (without invariants) we strengthen the hardness result to the claim that the problem is
at least as hard as the reachability problem for Vector Addition Systems, a problem for which even a primitive-recursive upper bound
is not known (see Section~\ref{sec:integers} for details).
These are the first complexity results for these problems, and
hopefully, another contribution is to motivate further research towards their theoretical understanding.
At the conclusion of this paper, further discussion of the significance of the results and the open problems will be given.

\section{Preliminaries}
\label{sec:prelim}

In this section we give basic definitions, regarding linear-constraint loops, linear ranking functions,
vector addition systems and inductive invariants.

\subsection{Loop representation}

We define the loop representation based on linear constraints, which is quite standard.

A \emph{single-path} linear-constraint loop with preconditions (\slcp for short) over
$n$ variables $x_1,\ldots,x_n$ has the form
\begin{equation*}
 C\vec{x} \le \vec{c};\quad
  \mathit{while}~(B\vec{x} \le
  \vec{b})~\mathit{do}~ A\begin{pmatrix}\vec{x}\phantom{'}\\
    \vec{x}'\end{pmatrix} \le \vec{a}
\end{equation*}
where $\vec{x}=(x_1,\ldots,x_n)^\trans$ and
$\vec{x}'=(x_1',\ldots,x_n')^\trans$ are column vectors, and for some
$p,q,r>0$, $C \in {\ints}^{r\times n}$, $B \in {\ints}^{p\times n}$, $A\in {\ints}^{q\times 2n}$,
$\vec{c} \in {\ints}^{r}$, $\vec{b}\in {\ints}^p$, $\vec{a}\in {\ints}^q$.
The constraint $C\vec{x} \le \vec{c}$ is called \emph{the precondition}, and specifies the initial states for
the computation of the loop.  The set of initial states is denoted by $\initials$.
The constraint $B\vec{x} \le \vec{b}$ is called \emph{the loop
  condition} (a.k.a. the loop guard) and the last constraint is
called \emph{the update}. 
The update is called \emph{deterministic} if, for a given $\vec x$
(satisfying the loop condition) there is at most one $\vec{x}'$
satisfying the update constraint. 

We say that there is a transition from a state $\vec{x}\in\rats^n$ to
a state $\vec{x}'\in\rats^n$, if $\vec{x}$ satisfies the condition and
$\vec{x}$ and $\vec{x}'$ satisfy the update.
A transition can be seen as a point $\trcv{\vec{x}}{\vec{x}'} \in \rats^{2n}$, where its first $n$
components correspond to $\vec{x}$ and its last $n$ components to
$\vec{x}'$. 
For convenience, we denote $\tr{\vec{x}}{\vec{x}'}$ by $\vec{x}''$.

The notions of \emph{computation of a loop} and termination are straight-forward. 
A computation must start at an initial state.
Note that when the
loop is non-deterministic, termination means that there exists no infinite computation from an initial
state.  A \emph{reachable state (transition)} is a state (respectively transition) that appears in some computation.

We say that the loop is interpreted \emph{over the rationals} if $\vec{x}$ and
$\vec{x}'$ range over $\rats^n$, and \emph{over the integers}
if they range over $\ints^n$. We also say that the loop is \emph{a rational (respectively, integer)} loop.
For  uniformity of notation, we use \states to denote the state space, without specifying its precise nature.

For purposes of complexity classification, we define the representation of the input to consist of the matrices
and vectors that specify the loop,
with numbers in binary notation.
We often consider a restricted problem, concerning a \emph{loop with partially-specified input}: that means that
the precondition is of the form $\bigwedge_{i=1}^k x_i=d_i$ for some variables $x_i$ and values $d_i$.
Thus the value of each variable is either specified precisely or left free.

\subsection{Ranking functions}

We now define linear ranking functions and
the decision problem  \linrfp, asking for the existence of a Linear Ranking Function for reachable states
(the $\rho$ reminds us of the lasso shape, and is also an initial of ``reachability").

An affine linear function $\rho: \rats^n \to \rats$ is 
of the form
$\rho(\vec{x}) = \vect{\rfcoeff}\cdot\vec{x} + \rfcoeff_0$ where
$\vect{\rfcoeff}\in\rats^n$ is a row vector and $\rfcoeff_0\in\rats$.
%
%

\bdfn
\label{def:linearrf}
Given a set $T\subseteq \rats^{2n}$, representing transitions,
we say that $\rho$ is a \emph{linear ranking function} (\lrf) for $T$
if the following
hold for every $\tr{\vec{x}}{\vec{x}'} \in T$:
\begin{align}
 \rho(\vec{x})  \ge 0  \,, \label{eq:lrf1}\\
 \rho(\vec{x})-\rho(\vec{x}')  \ge 1 \,. \label{eq:lrf2} 
\end{align}

We say that $\rho$ is a \lrf for a loop (with precondition) if its is a \lrf for the set of
\emph{reachable transitions} of this loop.
\edfn

\bdfn 
The decision problem \emph{Existence of a \lrf} (with precondition) is defined by

\begin{description}\setlength{\itemsep}{0pt}\setlength{\topsep}{0pt}\setlength{\itemindent}{0pt}
 \item[Instance:] an \slcp loop. 
 \item[Question:] does there exist a \lrf for this loop? 
\end{description}

\smallskip
\noindent
The decision problem is denoted by \linrfqp and \linrfzp for rational
and integer loops respectively. 
\edfn

\subsection{Invariants}

Consider a loop with initial states $\initials$ and transition set $\transitions$.
We define an \emph{invariant} of the loop to be a set $\inv\subseteq\states$ such that all reachable states
are in $\inv$.  We define an \emph{inductive invariant} (sometimes this is just called an invariant) to be a set
$\inv\subseteq\states$ satisfying the properties of
\begin{itemize}
\item  Initiation:
$\initials\subseteq \inv$;
\item Consecution:
if $\tr{\vec{x}}{\vec{x}'} \in \transitions$ then $\vec{x}\in\inv \Rightarrow \vec{x}'\in\inv$.
\end{itemize}
 Clearly, an inductive invariant does contain all reachable states. However, frequently, concentrating on inductive
 invariants makes the verification of an invariant possible---even if the precise set of reachable states could be 
uncomputable.  This depends on the kind of invariants one considers. For example, an often-used type of invariant
is \emph{convex polyhedra}~\cite{CousotHalbwachs:1978}.
Using a customary representation, e.g., by constraints, the invariant properties
are decidable by linear or integer programming (for linear-constraint loops). Another
natural class---for loops over the natural numbers---%
are \emph{downward-closed sets}:
 sets $\inv$ such that
$\vec{x}\le\vec{y}$, $\vec{y}\in\inv \Rightarrow$ $\vec{x}\in\inv$.
Due to Dickson's lemma, such sets are finitely representable 
as the downward-closure of a finite set
in the lattice $\nats_\omega^n$ (adding the element $\omega$ allows for unbounded sets in
$\nats$ to be represented). 
This makes them useful for analysing certain kinds of programs, notably vector addition systems~\cite{KM69,Hack79thesis}. We note
that, they constitute an elementary kind of disjunctive invariants---each disjunct
is of the form $0\le \vec{x}\le\vec{c}$ where $\vec{c}\in\nats_\omega^n$. 

Since with both of the above classes, verification of an invariant is effective, we call them \emph{effectively
inductive invariants}.
Our main interest lies in using the invariants to support ranking functions: this means that we look for a ranking function
not for the set of reachable transitions, but for the set $\{\tr{\vec{x}}{\vec{x}'} \mid \vec{x}\in\inv\}$, which may be larger,
but computable.

\section{Rational Loops with Preconditions}
\label{sec:rational}

Most of this section is dedicated to proving the next thoerem, from which we later derive the result on 
$\linrfqp$.

\begin{theorem} \label{thm:rational-reachability}
The following problem is PSPACE-hard: given a (deterministic) rational linear-constraint loop and an initial state,
does a specified variable ever get a positive value?
\end{theorem}

We prove this by reduction from the halting problem for Boolean programs, namely programs
that manipulate a finite number of $\{0,1\}$-valued variables,
$X_1,\dots,X_n$.  The program is a list of labeled instructions
$$1{:}I_1,\ldots,m{:}I_m,m{+}1{:}\stopsign$$ where 
each instruction ${I_k}$ is one of the following:
$$ 
incr(X_j) \mid decr(X_j) \mid
\mathit{if}~X_j~\mathit{then}~k_1~\mathit{else}~k_2
$$
with $1 \le k_1,k_2 \le m{+}1$ and $1 \le j \le n$.
A state is of the form $(k,\tuple{a_1,\ldots,a_n})$ which indicates
that Instruction $I_k$ is to be executed next,
and the current values of the variables are $X_1=a_1,\ldots,X_n=a_n$. In a valid state,
$1\le k\le m+1$ and all $a_i \in \{0,1\}$.
%
Any state in which $k=m+1$ is a halting state. For any other valid 
state ${(k,\tuple{a_1,\ldots,a_n})}$, the successor state is defined as follows.
\begin{itemize}
\item 
If $I_k$ is $incr(X_j)$, then $X_j$ is changed from 0 to 1; if it is already 1,
the program aborts.
Similarly, $decr(X_j)$ changes a 1 to a 0. In both cases, if execution does not abort,
  it proceeds at instruction $k+1$.

\item 
 If $I_k$ is
  ``$\mathit{if}~X_j~\mathit{then}~k_1~\mathit{else}~k_2$'', then the
  execution moves to instruction $k_1$ if $X_j$'s value is $1$, and to $k_2$ if
  it is $0$. The values of the variables do not change.
 \end{itemize}
The halting problem is whether the program reaches the halting label $m+1$ when
 started at the \emph{initial state}
$(1,\tuple{0,\dots,0})$ (note that aborting due to an invalid increment or decrement should
give a negative answer).

The class of deterministic Boolean programs captures PSPACE computability, 
and the halting problem for such programs is, therefore, PSPACE-complete (see, e.g., \cite{Jo:97}, which uses this
model up to non-essential differences).

Given a Boolean program $P_B$, 
we generate a corresponding \slcp loop $\trans({P_B})$ by translating the following program,
written in pseudo-code with assignments, into linear constraints.

\medskip
\begin{lstlisting}
while ( $0\le A_1 \le 1 \wedge\cdots\wedge 0\le A_m \le 1 \,\wedge\,
  0\le X_1\le 1 \wedge\cdots\wedge 0\le X_n\le 1$ ) do {
    $N_1$ := $0$; $N_2$ := $A_1$; $\dots$  $N_{m}$  := $A_{m-1}$; $N_{m+1}$ := $A_m+A_{m+1}$;
    $\trans(1{:}I_1)$
       $\vdots$
    $\trans(m{:}I_m)$
    $A_1$ := $N_1$;  $\dots$   $A_{m+1}$ := $N_{m+1}$
}
\end{lstlisting}
\par\noindent
Basically, $A_i$ represents the choice of instruction (the ``program counter"), and $N_i$ is a temporary
variable used for finding the next instruction (it is modified by jumps, as shown below).
$\trans({k{:}I_k})$ is a translation of the $k$th instruction, defined as follows (again, with a mix of assignments and assertions, for readability)

\begin{itemize} \setlength{\itemsep}{1ex plus0.2ex}
\item If ${I_k \equiv incr(X_j)}$, then $\trans({k{:}I_k})$ is $X_j :=
  X_j + A_k;$

\item If ${I_k \equiv decr(X_j)}$, then $\trans({k{:}I_k})$ is
$X_j := X_j - A_k;$

\item If $I_k \equiv \mathit{if}~X_j>0~\mathit{then}~k_1~\mathit{else}~k_2$, then
  $\trans(k{:}I_k)$ involves two dedicated variables, $T_k$ and $F_k$, as follows:
\par \(
\begin{array}{l}
0\le T_k  \le A_k; \\  
\phantom{0\le{}} T_k  \le X_j; \\  
0\le F_k \le  A_k; \\    
\phantom{0\le{}} F_k \le  1-X_j; \\    
T_k + F_k \ge A_k; \\
N_{k+1} := N_{k+1}-A_{k} ;\\
N_{k_1} := N_{k_1} + T_k;\\
N_{k_2} := N_{k_2} + F_k
\end{array}
\)
\end{itemize}

In the last part, the variables $T_k$ (respectively $F_k$) represent the choice of the ``true" branch (resp.~``false") of a
conditional branch instruction at label $k$.

Our precondition defines an initial state that corresponds to the initial state
of $P_B$. More precisely, in the initial state, all variables are set to 0, except $A_1 = 1$.
All auxiliary variables ($N_k$, $T_k$, $F_k$) are set to the appropriate values according to their
constraints, or to 0 if unconstrained. 
The essential arguments to complete the justification of the reduction are given by the following lemma.

\begin{lemma}
In every (rational-valued) state reachable from the initial state, it holds that
\begin{compactenum}
\item \label{correctness:01} all variables have values in $\{0,1\}$.
\item \label{correctness:T} $T_k= 1$ if and only if $A_k= 1$, instruction $k$ is a branch on $X_j$ and $X_j= 1$.
\item \label{correctness:F} $F_k= 1$ if and only if $A_k= 1$, instruction $k$ is a branch on $X_j$ and $X_j=0$.
\item \label{correctness:A} At most one variable $A_k= 1$. 
\item \label{correctness:halt} A state where all of $A_1,\dots,A_m$ are 0 is only reached when a 
transfer to label $m+1$ has been simulated. Only in such a state is $N_{m+1}=1$. 
\item \label{correctness:idle}
When a state where $N_{m+1}=1$ is reached, the program
idles in this state.
\end{compactenum}
\end{lemma}

\bprf
The proof requires induction on the number of transitions from the initial state. The initial state was
chosen to satisfy these properties. For the induction step, we first prove 
\eqref{correctness:T} and \eqref{correctness:F}, which follow quite easily (as the reader may check) from the assumption
that $A_k$ and $X_j$ are either 0 or 1 (which we have by the induction hypothesis). Given these facts, one can
check that for any state in which \eqref{correctness:01} and \eqref{correctness:A} hold, the variables $X_i$ remain
in $\{0,1\}$ (in fact, at most one of them is modified), proving that \eqref{correctness:01} holds in the next step.
For the variables $N_i$, since initially they are zero, it is
easy to see that it always holds that just one of them will be a 1 (using \eqref{correctness:A} and \eqref{correctness:halt}), which implies
\eqref{correctness:A} and \eqref{correctness:halt} for the next state. 
Finally, \eqref{correctness:idle} is easy to verify.
\eprf

Essentially, the lemma shows that the constraint loop simulates $P_B$ in lockstep (i.e., every
transition of $P_B$ is simulated by a transition of the loop), except that normal halting becomes an infinite
loop in a state where $N_{m+1}=1$.
Theorem~\ref{thm:rational-reachability} follows immediately.
By modifying  the constructed program slightly, we obtain

\bcor \label{cor:rational}
The \linrfqp problem is PSPACE-hard, even when restricted to deterministic loops.
\ecor

\bprf
We add another variable $Y$, initially unbounded, and the constraints:
$$  Y>0,\  Y' = Y-1+N_{m+1} \,.$$ 

It is easy to see that if the original Boolean program does \emph{not} halt, our \slcp loop will halt from the specified initial state,
and $Y$ is a ranking function. If the Boolean program \emph{does} halt, our \slcp loop does not, and therefore, has no
ranking function (of any kind).
\eprf

The fact that our loop either has the specified ranking function, or does not halt at all, is significant: it means that the existence
of any ``termination witness" (like \lrf) which can handle this loop (in particular, any witness which encompasses single-variable
\lrfs) will also be PSPACE-hard.
On the other hand, we can distinguish our problem from termination in the following sense.

\bcor \label{cor:rational-terminating}
The \linrfqp problem is PSPACE-hard even if restricted to deterministic loops that do terminate. 
\ecor

\bprf
We add another variable $R$, initially unbounded, and the constraints:
$$  R' =  R-1, \  Y' \le Y+R \,.$$ 
Now, the loop will always halt, since $R$ must eventually be negative and force $Y$ to decrease. But, when the Boolean program
halts, the loop can go through several iterations in which $Y$ does not decrease (as long as $R$ is still positive); therefore, $Y$ is still not
a ranking function.  Neither can we form a ranking function using other variables.
Specifically, we cannot use $R$, because it has no lower bound; and the rest of the variables
do not change in such iterations.
\eprf

\section{Integer Loops with Preconditions}
\label{sec:integers}

The constructions in this section are inspired by the simulation of Petri nets by \slcp loops, used for Theorem~6.1 in 
\cite{Ben-AmramGM:toplas2012}, which states that the termination problem for such loops is EXPSPACE-hard.
 The hardness result is based on Lipton's reduction from halting
of counter programs with exponential space;\footnote{The space complexity measure for counter programs
is the number of bits necessary to maintain the counters in binary notation.}
this was originally used by Lipton~\cite{Lipton76}
to prove hardness of some decision problems in Petri nets.
First, we give the necessary definitions.

\subsection{VAS and Petri nets}

A vector addition system  is a type of program which maintains $n$ counter variables (variables of
non-negative integer value), so that
a state $\vec{x}$ is a vector of
non-negative integers. A state-transition is of the form $\vec{x}' = \vec{x}+\vec{v}_i$, 
where $\vec{v}_i$ is chosen non-deterministically among
given \emph{displacement vectors}
 $\vec{v}_1,\dots,\vec{v}_k$, and subject to the constraint that all variables remain non-negative.
For purposes of complexity classification, we define the representation of a VAS as input to be the list of vectors,
with numbers in binary notation.  We denote the $j$th element of $\vec{v}_i$ by $\vec{v}_i[j]$.

A Petri net is a very similar model, and for convenience we present it here using the terminology of VAS.
Then, the difference lies in the definition of transitions: a possible transition is specified by \emph{two}
vectors, $\vec{v}^-_i$ and $\vec{v}^+_i$, both non-negative, and its effect is described by 
$\vec{x}' = \vec{x}-\vec{v}^-_i + \vec{v}^+_i$, provided $\vec{x}-\vec{v}^-_i$ is non-negative.
One may think of $\vec{v}^-_i$ as a requirement for the enabling of transition $i$.

\subsection{Lipton's reduction}

Let us first recall Lipton's reduction (a good reference is~\cite{Esparza98}).
Given an exponential-space counter program $P$,
the reduction constructs a Petri net $N_P$ that has the following
behavior when started at an appropriate initial state. $N_P$ has two
kinds of computations, \emph{successful} and \emph{failing}. Failing
computations are caused by taking non-deterministic branches which are
not the correct choice for simulating $P$. Failing computations
always halt. The (single) successful computation simulates $P$
faithfully. If (and only if) $P$ halts, the successful computation
reaches a state in which a particular flag, say $\textit{HALT}$, is
raised (that is, \textit{HALT} is a counter which is incremented for the first time from 0 to 1). 
This flag is never raised in failing computations.  Thus, the reduction proves hardness of a problem
which we may call \emph{eventual positivity}:

\bthm \label{thm:vas-pos-hard}
It is EXPSPACE-hard to decide, for a Petri net with a given initial state $\vec{x}_0$,
whether there is a reachable state in which $x_n > 0$.
\ethm

Note that this problem is a special case of \emph{coverability} (given $\vec{x}_0$ and
another vector $\vec{y}$, is there a reachable state $\vec{x}$ such that $\vec{x}\ge \vec{y}$?).
It is easy to adapt the reduction to also show hardness of state
reachability (is $\vec{y}$ reachable from $\vec{x}_0$?).

\subsection{Application to the LRF problem}
\label{sec:coverability2lrf}

By translating Petri nets to \slcp loops, we obtain a question on eventual positivity in
such loops with a partially-specified input. 
It is easy enough to transform this question to a question on the existence of a \lrf.
Thus we obtain

\begin{theorem} \label{thm:expspace-hard}
\linrfzp is EXPSPACE-hard for partially-specified input.
\end{theorem}

\bprf
Let a Petri net be given, having $n$ counter variables and $m$ displacement vectors,
along with an initial state $\vec{x}_0$.
We construct a \slcp loop
having variables $X_1,\ldots,X_n$, that represent the
counters, and flags $A_1,\ldots,A_m$, that
represent the choice of the next transition and change non-deterministically.
The loop guard is $X_1\ge 0\wedge\cdots\wedge X_n\ge 0$. 
The initial state for our loop is the given initial state (for the $X_i$) and zeros for the $A_i$.
The transition relation of the loop implements the Petri-net transitions in a straight-forward way,
specifically, it is the conjunction of the following three conjunctions
\begin{eqnarray*}
\Delta &\equiv&  \bigwedge_{k=1}^m (A_k' \ge 0) \wedge (A_1'+....+A_m' = 1) \\
\Psi   &\equiv& \bigwedge_{i=1}^n (X_i \ge \sum_{k=1}^m \vec{v}_k[i] \cdot A_k') \\
\Phi   &\equiv& \bigwedge_{i=1}^n (X_i' = X_i - \sum_{k=1}^m \vec{v}^-_k[i] \cdot A_k'  
                                                                                       + \sum_{k=1}^m \vec{v}^+_k[i] \cdot A_k')
\end{eqnarray*}
where $\Delta$ ensures that one and only one $A_k'$ will be a 1, $\Psi$ ensures that the transition
chosen is enabled, and $\Phi$ implements the effect of the transition.

 To reduce to the  \linrfzp problem, we add another variable $Y$, and the constraints:
$$  Y>0,\  Y'\le Y-1+X_n  .$$ 
In addition, we add a new transition to our VAS (and encode it in our loop);
the new transition is enabled when  $X_n$ is positive, and does not modify the state (so it loops forever).

It is easy to see now
that if the original counter program does \emph{not} halt, our \slcp loop will halt from the specified initial state,
because variable $Y$ will hit its lower bound; in fact,
$Y$ is a ranking function. If the original counter program \emph{does} halt, our \slcp loop does not, and therefore, has no
ranking function (of any kind).

We conclude that determining if the constructed loop has a linear ranking function is as hard as deciding whether the counter
machine that the Petri net simulates halts, that is, EXPSPACE-hard.
\eprf

Note that a reduction from counter programs with unbounded counter values would have proved undecidability. 
Unfortunately, such a reduction has not yet been found. The reduction from VAS suceedes, essentially, because it is a kind
of counter program in which a transition cannot be conditioned on a zero-test.

\subsection{A reduction from Reachability}

In the Reachability problem for Petri nets/VAS, we are given an initial state $\vec{s}$ and are asked whether 
a given target state $\vec{t}$ is reachable from $\vec{s}$. We can also reduce to \linrfzp from the reachability
problem. This observation may be of interest since the latter problem is generally presumed to be harder than
coverability, which provided our EXPSPACE lower bound \cite{demri-esslli2010,Esparza94,Leroux:2011:VAS}
(at least, it is certain that reachability too is EXPSPACE-hard, so the same lower bound follows. Hence, in terms of the
resulting lower bound, the next theorem
supersedes the previous one. However, the previous reduction is not entirely redundant as it is useful for a proof to be
given later in Section~\ref{sec:invariants}).

Intuitively, the reduction operates as follows: a program simulates the VAS and tries to check if the target vector
$\vec{t}$ is reached. When this happens, it results in an infinite execution.

\begin{theorem}
There is polynomial-time reduction of the VAS reachability problem to 
\linrfzp, with a partially-specified initial state.
\end{theorem}

\bprf
Let a VAS (of dimension $n$, and with $m$ displacement vectors),
and the vectors $\vec{s}$, $\vec{t}$ be given.
We assume (with no loss of generality) that the VAS is designed so that a computation from $\vec{s}$ will never
reach the zero vector. 

 We construct a constraint loop simulating it as follows.
The loop has variables $X_1,\dots,X_n,X_{n+1}$ and $A_1,\dots,A_{m+2}$. The guard is
\[
 X_1\ge 0\wedge\cdots\wedge X_{n+1}\ge 0 \wedge A_1\ge 0 \wedge\cdots\wedge A_{m+2}\ge0 \wedge
 \sum_i A_i = 1
 \]
 and the update is the conjunction of the following constraints, 
 \begin{align}
\label{eq:update}
  &X_j' = X_j + ( \sum_{i=1}^m \vec{v}_i[j] \cdot A_i )  - \vec{t}[j]\cdot A_{m+1}  &&\text{for $j=1,\dots,n$}, \\
\label{eq:choose}
  &A'_1\ge 0 \wedge\cdots\wedge A'_{m+2}\ge 0 \wedge \sum_i A'_i = 1, \\
\label{eq:decrease}
  &X_{n+1}' = X_{n+1} - (\sum_{j=1}^n X_j), \\
\label{eq:stay}
  &A'_{m+2} \ge A_{m+1}+A_{m+2} \,.
  \end{align}
The initial state for our loop is just the given initial state $\vec{s}$ (for the $X_i$) and unspecified for the $A_i$.

\emph{Explanation}:  As before, the variables $X_1,\dots,X_n$ and $A_1,\dots,A_m$ are used to simulate the VAS.
The $X$'s represent the state vector $\vec{x}$, and
$A$'s are flags which change non-deterministically to indicate the next transition.
This simulation goes on as long as $A_{m+1}$ or $A_{m+2}$ have not turned on, and as long as it does go on,
$X_{n+1}$ descends, by \eqref{eq:decrease}.  If $A_{m+1}$ turns on, the target vector $\vec{t}$ is substracted
from $(X_1\dots X_n)$, so such a transition is only enabled if this vector $\vec{x}$ is at least as large as $\vec{t}$.
Suppose that this happens. Then by \eqref{eq:stay}, later transitions are forced to have $A_{m+2}=1$,
so they do not simulate the loop any longer, and the $X$'s do not change, except for $X_{n+1}$, which continues
to decrease if and only if $\vec{x}$ at the start of this phase was \emph{not equal} to $\vec{t}$.

Hence, if $\vec{t}$ is reachable from $\vec{s}$, it is possible to run into a non-terminating computation where
nothing decreases, and the loop has no \lrf. Otherwise, $X_{n+1}$ keeps decreasing, even when 
$A_{m+2}=1$, so it constitutes a \lrf.

We should note that it is possible to turn $A_{m+2}$ on without passing through $A_{m+1}=1$, and in this case 
$X_{n+1}$ keeps decreasing regardless of the reachability question, so our reduction remains correct.
\eprf

\subsection{Ranking versus termination}

As in Section~\ref{sec:rational}, we can see that our reduction yields a loop which 
either has the specified ranking function, or does not halt at all, which means that the existence
of any ``termination witness" (like \lrf) which can handle the loop (in particular, any witness which encompasses single-variable
\lrfs) will also be PSPACE-hard.
On the other hand, we can show (using the same trick as in Corollary~\ref{cor:rational-terminating}) that
the \linrfzp problem is EXPSPACE-hard even if restricted to loops that do terminate. 

\subsection{Deterministic loops}

Both of the above hardness results also hold for deterministic constraint loops. In order to do that, we need to 
``determinize" the loop constructed in the reduction. The technique is from~\cite{Ben-AmramGM:toplas2012},
and consists of adding an uninitialized variable, whose value is used as an ``oracle," to guide the non-deterministic choices.
In the case that there is a computation which makes the value of $X_n$ positive (in our first reduction) or the value of
$\sum_{j=1}^n X_j$ zero (for the second), there will be a value for the oracle variable that guides the computation to this state.
See~\cite[Sect.~6.1]{Ben-AmramGM:toplas2012} for more details.

\section{Hardness for Invariant-Supported LRFs}
\label{sec:invariants}

In this section we turn to the problem of invariant-supported LRFs, namely the decision problem defined as follows:

\begin{description}\setlength{\itemsep}{0pt}\setlength{\topsep}{0pt}\setlength{\itemindent}{0pt}
 \item[Instance:] an \slcp loop. 
 \item[Question:] does there exist an inductive invariant $\inv$ (of a particular class)  for this loop, such that
 there is a \lrf for $\{\tr{\vec{x}}{\vec{x}'} \mid \vec{x}\in\inv\}$?
\end{description}

\noindent
We give two hardness results, depending on the type of invariant: PSPACE-hardness for convex polyhedra
and EXPSPACE-hardness for downward-closed sets over $\nats^n$. Both are derived from the constructions of earlier sections,
by noticing that if there is a ranking function, there is an invariant to support it. Both address integer loops only.

\bthm
For deterministic integer \slcp loops, 
deciding whether convex polyhedral invariant exists which supports a \lrf for the loop is PSPACE-hard.
\ethm

\bprf
We use the reduction from halting of Boolean programs (Section~\ref{sec:rational}). We claim that when there is
a \lrf (which would consist of the variable $Y$ as shown in the proof of Corollary~\ref{cor:rational}), then there
is a convex polyhedral invariant supporting it.  Indeed, assume that the Boolean program does not halt.
Let $\poly{R}$ be the set of reachable states of the \slcp loop constructed. In all these states, the variables are
0-1 valued (except for $Y$, which is unbounded). Note that the convex hull of a set $V$ of 0-1 vectors includes no 
other integer vectors besides $V$. Thus, as we are considering an integer loop,
the convex hull of $\poly{R}$ represents $\poly{R}$ precisely, and constitutes an invariant (the set of reachable states
is clearly an inductive invariant) which supports the \lrf.
\eprf

\bthm
For deterministic \slcp loops over the natural numbers,
deciding whether a downward-closed invariant exists which supports a \lrf for the loop is EXPSPACE-hard.
\ethm

\bprf
We reuse the reduction from the Eventual Positivity problem of Petri nets (Theorem~\ref{thm:expspace-hard}). We claim that when there is
a \lrf (which would consist of the variable $Y$ as shown in the proof), then there
is a downward-closed invariant supporting it.  
Let $\poly{R}$ be the set of reachable states form the given initial states, and
consider its downward-closure $\check{\poly{R}}$. It clearly contains the initial states, and it is easy enough to verify that it is closed
under the transition relation (recall that $\poly{R}$ is closed, by definition). Thus $\check{\poly{R}}$
constitutes an invariant; it supports the \lrf $Y$ because $X_n$ is zero in all these states.
\eprf

As in previous sections, both of the above hardness results are also valid when restricting to programs that do terminate.

\section{Related work}

Termination analysis has been the subject of many papers (too many to list here), but, in addition to works already mentioned,
the following works seem closely related.

There are some tools which, like \textsc{Terminator}, use a counterexample-directed approach which naturally calls for the
analysis of lasso programs. Examples include~\cite{PR2007armc,harris2011alternation,DBLP:conf/tacas/CookSZ13}.

Bagnara et al.~\cite{DBLP:journals/iandc/BagnaraMPZ12} give a clear exposition on the computation of universal
ranking functions, comparing \cite{DBLP:conf/vmcai/PodelskiR04} with previous solutions~\cite{SohnVanGelder:91,MS:08} that
use essentially the same approach.
Recently, several works addressed the generation of more complex termination proofs, in particular, involving 
lexicographically-decreasing tuples of linear functions. The universal problem for linear-constraint loops is analysed in
\cite{Ben-AmramG13jv}, while preconditions have been taken into account in some works: Bradley et al.~\cite{BMS:LexLinear,BMS:Polyranking}
search for supporting invariant using constraint solving (for multi-path loops). Alias et al.~\cite{ADFG:2010} handle control-flow graphs
of any form, but require precomputed invariants.   Brockschmidt et al.~\cite{DBLP:conf/cav/BrockschmidtCF13} use an iterative method
in which invariant generation is guided by the needs of the termination prover. They use a separate safety checker to provide them,
while Larraz et al.~\cite{LarrazORR13} use the constraint-solving approach to find supporting invariants together with the
ranking functions, but using iterative improvement as in the latter work.

Regarding the computation of \emph{termination preconditions}, Bozga et al.~\cite{BIKtacas2012jv} show
that for loops specified by
\emph{octagonal relations} a precondition for \emph{termination} can be computed in polynomial
time.  But the proof does not necessarily produce a linear ranking function.

\section{Concluding Remarks}

We have established lower bounds (that is, hardness results) on the complexity of the linear-ranking
problem with preconditions, first in its general form and then when restricted to \lrfs supported by
two forms of effectively inductive invariants.
In fact, our lower bounds hold for the linear-ranking function verification problem:

\begin{description}\setlength{\itemsep}{0pt}\setlength{\topsep}{0pt}\setlength{\itemindent}{0pt}
 \item[Instance:]  an \slcp loop and an affine-linear function $f$. 
 \item[Question:] is $f$ a \lrf for this loop? 
\end{description}

 Moreover, the lower bounds hold for a simple kind of precondition, namely a partially-specified input.
Even for this case, we do not have upper bounds, and obtaining them seems extremely difficult.  
We still have no answer to the following intriguing questions: \emph{Is any of the decision problem studied
here any easier than termination for the corresponding class of loops? And are they equivalent to reachability?}

An interesting open problem results from restricting the update, say to \emph{affine linear}: 
$\vec{x}' = A'\vec{x} + \vec{a}'$.    We note that the 
notorious \emph{positivity problem for linear recurrence sequences} \cite{OuaknineWorrell:SODA2014}
translates easily to \lrf verification---to check whether $x_1$ is always positive, add a variable $x_{n+1}$ with $x'_{n+1} = x_{n+1}-x_1$,
put $x_{n+1}\ge 0$ in the guard and ask whether $f(\vec x) = x_{n+1}$ is a \lrf.
It would be interesting to know whether a reduction to \lrf \emph{existence} can also be found.

An interesting direction for further research may be to find out the implications of fixing the number of variables.
For our VAS problem it is known to make the problem solvable in polynomial space \cite[Corollary 3.4.5]{demri-esslli2010}. Our lower bounds do not give any significant result in this case
(and clearly, the problem does get easier for sufficiently small $n$: at least for $n=1$ it does!).

Another, very natural, idea is to restrict the invariants to a fixed (or polynomial) number of conjuncts or disjuncts.
In fact, all the works using the constraint-solving approach are based
on such a restriction.  But is the problem tractable now? There is again an intriguing  lack of results.
For conjunctions of a given number of linear constraints,
Bradely, Manna and Sipma~\cite{BMS:LexLinear} show decidability in exponential time, but only over the reals
(it is not clear whether results would be different over the rationals). 
Heizmann et al.~\cite{HeizmannHLP:ATVA2013} show a polynomial-time procedure for loops over
the reals or rationals, when the invariant is a single half-space; and they prove completeness only under an additional restriction.
So there is a long way ahead.


\end{document}